\documentclass[aps,prb,twocolumn,preprintnumbers,amsmath,amssymb,superscriptaddress]{revtex4-2}
\usepackage{amsmath,graphicx}
\usepackage[utf8]{inputenc}
\usepackage[T1]{fontenc}
\usepackage{xcolor}
\usepackage{textcomp}
\usepackage{bm}

\usepackage{siunitx}
\usepackage{physics}
\usepackage{amsmath}
\usepackage{tikz}
\usepackage{mathdots}
\usepackage{yhmath}
\usepackage{cancel}
\usepackage{color}
\usepackage{array}
\usepackage{multirow}
\usepackage{amssymb}
\usepackage{gensymb}
\usepackage{tabularx}
\usepackage{extarrows}
\usepackage{booktabs}
\usetikzlibrary{fadings}
\usetikzlibrary{patterns}
\usetikzlibrary{shadows.blur}
\usetikzlibrary{shapes}

\usepackage{color}
\definecolor{LinkColor}{rgb}{0.75,0.0,0.2}

\usepackage{hyperref}
\hypersetup{
	pdfauthor={good guys},
	pdftitle={good title},
	colorlinks=true,
	citecolor=LinkColor,
	linkcolor=LinkColor,
	urlcolor=LinkColor,
}

\usepackage{listings}
\definecolor{lightgray}{gray}{1}

\begin{document}
\title{Noise-induced phase transitions in hybrid quantum circuits}

\author{Shuo Liu}
\affiliation{Institute for Advanced Study, Tsinghua University, Beijing 100084, China}
\author{Ming-Rui Li}
\affiliation{Institute for Advanced Study, Tsinghua University, Beijing 100084, China}
\author{Shi-Xin Zhang}
\email{shixinzhang@iphy.ac.cn}
\affiliation{Institute of Physics, Chinese Academy of Sciences, Beijing 100190, China}
\author{Shao-Kai Jian}
\email{sjian@tulane.edu}
\affiliation{Department of Physics and Engineering Physics, Tulane University, New Orleans, Louisiana, 70118, USA}
\author{Hong Yao}
\email{yaohong@tsinghua.edu.cn}
\affiliation{Institute for Advanced Study, Tsinghua University, Beijing 100084, China}
\date{\today}

\begin{abstract}
The presence of quantum noises inherent to real physical systems can strongly impact the physics in hybrid quantum circuits with local random unitaries and mid-circuit measurements. The quantum noises with
a size-independent occurring probability can lead to the disappearance of a measurement-induced entanglement phase transition and the emergence of a single area-law phase. 
In this work, we investigate the effects of quantum noises with size-dependent probabilities $q=p/L^{\alpha}$ where $\alpha$ represents the scaling exponent. 
We have identified a noise-induced entanglement phase transition from a volume law to a power (area) law in the presence (absence) of measurements as $p$ increases when $\alpha=1$. 
With the help of an effective statistical model, we reveal that the phase transition is of first-order arising from the competition between two types of spin configurations and shares the same analytical understanding as the noise-induced coding transition. This unified picture further deepens the understanding of the connection between entanglement behavior and the capacity of information protection. When $\alpha \neq 1$, one spin configuration always dominates regardless of $p$ and thus the phase transition disappears. Moreover, we highlight the difference between the effects of size-dependent bulk noise and boundary noises. We validate our analytical predictions with extensive numerical results from stabilizer circuit simulations. 
\end{abstract}

\maketitle

\section{Introduction}
Measurement-induced phase transitions (MIPTs)~\cite{PhysRevB.98.205136, PhysRevB.100.134306, PhysRevX.9.031009} have recently attracted significant attention and have been investigated in various setups~\cite{PhysRevLett.126.060501, PRXQuantum.2.040319, PhysRevX.12.011045, PhysRevX.10.041020, PhysRevLett.125.030505, PhysRevB.99.224307, PhysRevB.100.064204, PhysRevB.101.104301, PhysRevB.103.174309, PhysRevB.103.104306, PhysRevB.101.104302, PhysRevB.107.214203, hokeMeasurementinducedEntanglementTeleportation2023, PhysRevB.105.104306, PhysRevB.102.014315, PhysRevB.106.214316, PhysRevB.106.144311, PRXQuantum.4.030333, PhysRevX.12.041002, PhysRevB.107.014308, PhysRevLett.129.120604, Han2023quantum, PhysRevLett.127.140601, PhysRevB.106.224305, MIPT_SYK_2, NonlocalMIPT_Qi,PRXQuantum.2.010352, NonlocalMIPT_Quantum, PhysRevB.104.094304, PhysRevLett.128.010605, PhysRevLett.128.010604, PhysRevResearch.4.013174, PhysRevLett.128.010603, 10.21468/SciPostPhysCore.5.2.023, Zhang2022universal, PhysRevB.108.L041103, ravindranath2023free, PhysRevB.109.L020304, PhysRevLett.130.120402, PhysRevB.108.054307, Biella_2021, PhysRevB.103.224210, PhysRevB.105.L241114,PRXQuantum.5.020304, akhtar2023measurement, PhysRevLett.131.020401, PhysRevB.107.094309}. 
These studies have revealed that the entanglement within a system undergoes a transition from a volume law to an area law as the measurement probability increases. 
However, in real experimental quantum systems, coupling to the environment unavoidably introduces quantum noises. 
In terms of the effective statistical model for random quantum circuits, the quantum noises can be treated as symmetry-breaking fields that result in the disappearance of the entanglement phase transition and a single area-law entanglement phase regardless of the measurement probability~\cite{BAO2021168618, Noise_bulk, jian2021quantum, PhysRevB.108.104310, PhysRevB.107.014307, PhysRevB.108.104310, PhysRevB.107.L201113, PhysRevLett.132.240402, Zhang_2022}. 

The MIPT from a power law phase to an area law phase with fixed quantum noises at the spatial boundaries has been investigated~\cite{PhysRevLett.129.080501}, which can be regarded as a special case of quantum noises with size-dependent probabilities $q=2/L$ where $L$ is the system size. Additionally, the effects of quantum noises or $T$ gates in the bulk with size-dependent probabilities $q=p/L$ have been explored in the context of random circuit sampling~\cite{NIPT1_arxiv, NIPT2_arxiv, PhysRevA.109.042414} and non-stabilizerness transition~\cite{bejan2023dynamical, fux2023entanglement}. However, the investigation of the entanglement phase transition in the MIPT setup with bulk quantum noises of size-dependent probability is lacking. Moreover, the entanglement structures and critical behaviors associated with quantum noises of size-dependent probabilities, as well as the influence of different choices of scaling exponents $\alpha$ are also worth studying.

The entanglement structure and information protection capacity are closely related~\cite {PhysRevLett.125.030505, PhysRevLett.111.127205, PhysRevX.7.031016, PhysRevX.11.011030, QI_NP, PhysRevX.10.041020}. 
From the perspective of information protection, a spatial boundary and a temporal boundary noise-induced coding transition both occur~\cite{Coding_Vijay, PhysRevLett.132.140401}. Below a finite critical probability of boundary noise, the encoded information can be protected after a hybrid evolution of time $O(L)$. 
On the contrary, if the probability of boundary noise exceeds this critical value, the information will be destroyed by quantum noises. 
A similar noise-induced coding transition is anticipated in the presence of bulk quantum noise with probability $q=p/L$, but the differences in information protection between boundary and bulk noises have not been explored before.
Furthermore, a comprehensive theoretical understanding of the connections between noise-induced entanglement and coding transitions is highly desired.

In this work, we investigate the entanglement phase transition in the presence of quantum noises with size-dependent probability in a MIPT setup. We have identified a more general entanglement phase diagram, as shown in Fig.~\ref{fig:setup} (c), where the x-axis and y-axis represent measurement probability $p_{m}$ and noise probability prefactor $p$, respectively. Besides the original MIPT occurring at $p_{m}=p_{m}^{c}$ and $p=0$, we identify a noise-induced entanglement phase transition from a volume law phase to a power (area) law phase when $0< p_{m} < p_{m}^{c} \sim 0.3$~\cite{Noise_bulk}  $(p_{m}=0)$, denoted by the black solid line in Fig.~\ref{fig:setup} (c). 
Via mapping to the classical spin model, the entanglement phase transition can be understood as the competition between two types of spin configurations, and the power law entanglement is attributed to the Kardar-Parisi-Zhang (KPZ) fluctuation~\cite{PhysRevLett.56.889, PhysRevLett.55.2923, PhysRevLett.55.2924, PhysRevA.39.3053, PhysRevLett.129.080501, Gueudré_2012,barraquandHalfSpaceStationaryKardar2020} with an effective length scale $L_{\text{eff}} \sim L/p$~\cite{PhysRevB.107.L201113, PhysRevLett.132.240402}.

Besides, we have also investigated the coding transition in the presence of size-dependent quantum noises. 
The analytical picture of the coding transition is distinct from that presented in Ref. \cite{Coding_Vijay}, where quantum noises are only applied on one spatial boundary. 
Theoretically, we reveal that the noise-induced entanglement phase transition and coding transition can be understood within the same framework, further establishing the connection between entanglement structure and information protection capacity. 
These two transitions are both first-order transitions at the same critical point $p_c$ and with the same critical exponent $\nu=2$~\cite{Coding_Vijay}. 
To validate our theoretical findings, we have conducted stabilizer circuit simulations, providing compelling evidence for the existence and universal behavior of noise-induced entanglement phase and coding transitions. 

We remark that the choice of the scaling exponent $\alpha$ for quantum noises is crucial for noise-induced phase transitions. 
Previous studies focused on the disappearance of the entanglement phase transition in the presence of bulk quantum noises~\cite{BAO2021168618, Noise_bulk, jian2021quantum, PhysRevB.108.104310, PhysRevB.107.014307, PhysRevB.108.104310, PhysRevB.107.L201113, PhysRevLett.132.240402}, corresponding to $\alpha=0$. 
In this work, we demonstrate that noise-induced phase transitions only occur at $\alpha=1$. 
In terms of the effective statistical model, there is only one single dominant spin configuration for $\alpha\neq 1$ with no competition between different configurations. 

The remainder of the paper is outlined as follows. In Sec.~\ref{sec:setup}, we introduce the setups of noise-induced entanglement phase transition and coding transition and the associated observables. In Sec.~\ref{sec:statistical}, we introduce the unified theoretical understanding of these two different noise-induced phase transitions with the help of the effective statistical model. In Sec.~\ref{sec:numerical}, we show the numerical results supporting the theoretical understanding. In Sec.~\ref{sec:distinction}, we discuss the distinctions between bulk noises and boundary noises. Finally, the conclusion and discussions follow in Sec.~\ref{sec:conclusion}. The additional numerical results are shown in the Appendix.

\section{Setup and observables}
\label{sec:setup}

\begin{figure}[ht]
\centering
\includegraphics[width=0.46\textwidth, keepaspectratio]{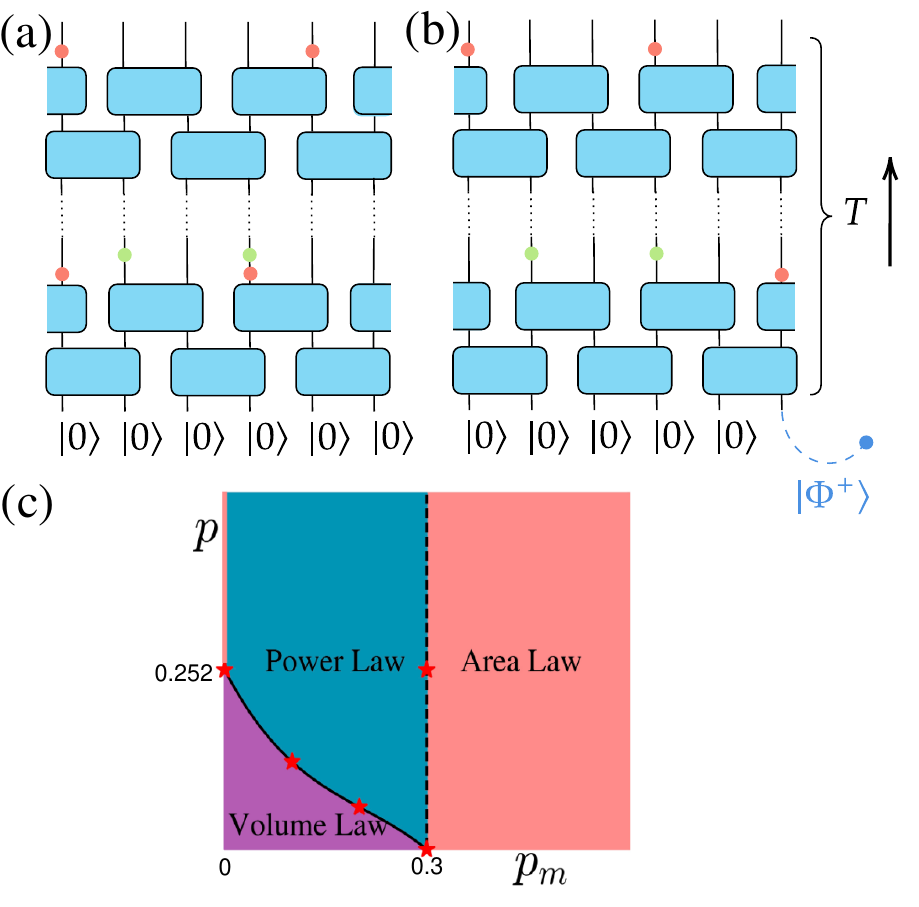}
\caption{Circuit setups with $6$ qudits for (a) entanglement phase transition and (b) coding transition. 
The red and green circles represent the quantum channels and projective measurements, respectively. 
In (b), a qudit is maximally entangled with a reference qudit by creating a Bell pair $\vert \Phi^{+} \rangle$ to encode one qudit information. 
(c) Phase diagram of the entanglement phase transition with $T=4L$. 
Red stars represent the critical points identified from numerical results. 
The black solid (dashed) curve denotes the noise (measurement)-induced phase transition.
}
\label{fig:setup}
\end{figure}

To investigate the noise-induced entanglement phase transition, we consider a one-dimensional system composed of $L$ $d$-qudits with initial state $\vert 0 \rangle^{\otimes L}$, as illustrated in Fig. \ref{fig:setup} (a). 
At each discrete time step, a layer of random two-qudit unitary gates arranged in a brick-wall structure with periodic boundary conditions (PBC) is applied. Then the projective measurements and quantum noises act on each qudit with probability $p_{m}$ and $q=p/L^{\alpha}$, respectively. 
The hybrid evolution time is $T=4L$ unless otherwise specified.

To quantify the entanglement for the final mixed state~\cite{entropy_fail_1, entropy_fail_2}, we employ the logarithmic entanglement negativity~\cite{Negativity_PhysRevA02, Negativity_PhysRevLett05, Negativity_PhysRevLett12, Negativity_Calabrese, Negativity_PhysRevB19, Negativity_PhysRevLett20, Negativity_PhysRevB20, Negativity_PhysRevLett20_Wu, Negativity_PRXQuantum21, Negativity_Shapourian} between the left ($A$) and right ($B$) half chain of the final state
\begin{eqnarray}
   E_{N} = \log \vert \vert \rho^{T_{B}} \vert \vert_{1},
\end{eqnarray}
where $\rho^{T_{B}}$ is the partial transpose of $\rho$ in subsystem $B$ and $\vert \vert \cdot \vert \vert_{1}$ denotes the trace norm. We also calculate the mutual information 
\begin{eqnarray}
    I_{A:B}=S_{A}+S_{B}-S_{AB},
\end{eqnarray}
where $S$ is the von Neumann entropy. Mutual information gives qualitatively similar scaling to $E_{N}$ and provides a more intuitive understanding within the framework of the statistical model.

The setup for the noise-induced coding transition is similar. 
The main difference is that one qudit of the system is maximally entangled with a reference qudit to encode one qudit of information at the initial state, as shown in Fig. \ref{fig:setup} (b). 
The choice of the qudit is arbitrary due to PBC. To quantify the information that remained in the system in the presence of quantum noises, we measure the mutual information $I_{AB:R}=S_{AB}+S_{R}-S_{AB \cup R}$ between the system and the reference qudit. $I_{AB:R}=2$ ($0$) means that the encoded information is perfectly protected (destroyed). To compare the entanglement phase transition and the coding transition, we set their evolution times $T$ to be equal.

\section{Statistical model}
\label{sec:statistical}

In this section, we present the theoretical understanding of noise-induced phase transitions via the mapping between the hybrid quantum circuit and the effective statistical model.
We focus on mutual information for simplicity. 
Please refer to Refs.~\cite{PhysRevB.107.L201113, PhysRevLett.132.240402} for more details of the effective statistical model and the analysis of $E_{N}$. 

\subsection{Mapping between the quantum circuit and the statistical model}

The $(1+1)$D quantum circuit can be mapped to a 2D classical spin model after averaging independent two-qudit random Haar gates. For simplicity, we begin with the most basic setup without any quantum noise or projective measurement and defer the discussions of cases with quantum noises or projective measurements to the subsequent sections.

    \begin{figure*}[ht]
    \centering
    \includegraphics[width=0.95\textwidth, keepaspectratio]{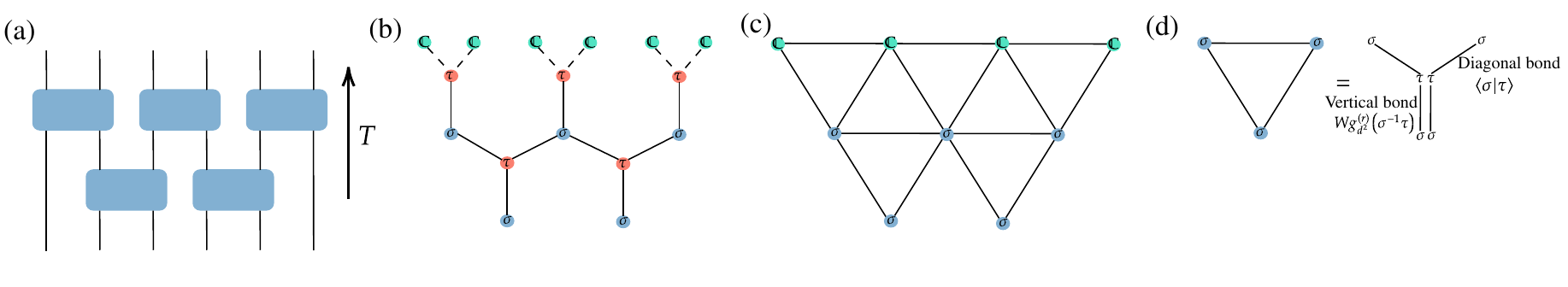}
    \caption{(a) shows the quantum circuit without quantum noise and projective measurement. The random two-qudit unitary gates (blue rectangles) are arranged in a brick-wall structure. (b) shows the statistical model with the degrees of freedom formed by the permutation-valued spins $\sigma$ and $\tau$. For the statistical model corresponding to $S_{AB}$, the top boundary is fixed by adding an additional layer of spins $\mathbb{C}$. (c) we integrate out spins $\tau$ to obtain the positive three-body weights of the downward triangles. The total weight is the product of the weights of downward triangles. (d) shows the three-body weight of a downward triangle.}
    \label{fig:mapping_main}
    \end{figure*}

The quantum circuit at a given trajectory consists of random two-qudit unitary gates arranged in a brick-wall structure as shown in Fig.~\ref{fig:mapping_main} (a). The density matrix $\rho$ after evolution time $T$ is
\begin{eqnarray}
    \rho(T) = (\prod_{t=1}^{T} \tilde{U}_{t}) \rho_{0} (\prod_{t=1}^{T} \tilde{U}_{t})^{\dagger},
\end{eqnarray}
where $\rho_{0}$ represents the density matrix of the initial state, and $\tilde{U}_{t}$ is the unitary evolution of discrete time step $t$ which is given by
\begin{eqnarray}
    \tilde{U}_{t}= \prod_{i=0}^{\frac{L-2}{2}} U_{t, (2i+2, 2i+3)}  \prod_{i=0}^{\frac{L-2}{2}} U_{t, (2i+1, 2i+2)}, 
\end{eqnarray}
where each two-qudit unitary gate is independently and randomly drawn from the Haar measure. We choose PBC and thus $L+i \equiv i$.
To obtain analytical results, we first express $| \rho(T) \rangle$ in an $r$-fold replicated Hilbert space
\begin{eqnarray}
    \vert \rho(T) \rangle^{\otimes r} &=& \prod_{t=1}^{T} \left[ \tilde{U}_{t} \otimes \tilde{U}_{t}^{*}  \right]^{\otimes r} \vert \rho_{0} \rangle^{\otimes r} \\ \nonumber 
    &=& \prod_{t=1}^{T} ( \prod_{i=0}^{\frac{L-2}{2}} (U_{t, (2i+2, 2i+3)} \otimes U^{*}_{t, (2i+2, 2i+3)})^{\otimes r}  \\ \nonumber
    && \prod_{i=0}^{\frac{L-2}{2}} (U_{t, (2i+1, 2i+2)} \otimes U^{*}_{t, (2i+1, 2i+2)})^{\otimes r}  ) \vert \rho_{0} \rangle^{\otimes r}.
\end{eqnarray}
Each random two-qudit unitary gate $U_{t,(i,j)}$ can be averaged independently~\cite{PhysRevB.101.104301, PhysRevB.101.104302, PhysRevX.7.031016, PhysRevLett.129.080501, PhysRevB.99.174205, PhysRevX.12.041002, collins2003moments, collinsIntegrationRespectHaar2006, PhysRevX.8.021014}:
\begin{eqnarray}
    && \mathbb{E}_{\mathcal{U}}(U_{t,(i,j)} \otimes U_{t,(i,j)}^{*})^{\otimes r}  \\ \nonumber
    &=& \sum_{\sigma, \tau \in S_{r}} \text{Wg}_{d^2}^{(r)}(\sigma \tau^{-1}) \vert \tau \tau \rangle \langle \sigma \sigma \vert_{ij},
\end{eqnarray}
where $d$ is the local Hilbert space dimension of qudit, and $\sigma, \tau$ are permutation spins in the permutation group $S_{r}$ of dimension $r$. We showcase the exact expressions of permutation spins when $r=2$ (two-copy). There are two types of spins: one is the identity permutation spin $\mathbb{I} = \sum_{i,j=0}^{d-1} \vert i_{\rm{k}, 1}i_{\rm{b}, 1}j_{\rm{k}, 2}j_{\rm{b}, 2} \rangle$ where $i \ (j)$ represents the computational basis of a qudit and the index k (b) represents ket (bra) for the first or second copy, the other is the swap permutation spin $C=\sum_{i,j=0}^{d-1} \vert i_{\rm{k}, 1}j_{\rm{b}, 1}j_{\rm{k}, 2}i_{\rm{b}, 2} \rangle$. $\text{Wg}_{d^{2}}^{(r)}$ is the Weingarten function with an asymptotic expansion for large $d$~\cite{PhysRevB.99.174205, collinsIntegrationRespectHaar2006}:
\begin{eqnarray}
    \text{Wg}_{d^{2}}^{(r)} (\sigma)  = \frac{1}{d^{2r}} \left[ \frac{\text{Moeb}(\sigma)}{d^{2\vert\sigma\vert}} + \mathcal{O}(d^{-2\vert \sigma \vert -4})\right],
\end{eqnarray}
where $\vert \sigma \vert$ is the number of transpositions required to construct $\sigma$ from the identity permutation spin $\mathbb{I}$.

Via regarding the permutation spins as the degrees of freedom, we can transform the quantum circuit into a classical statistical model. The partition function $Z$ of this effective statistical model is the summation of the total weights of various spin configurations, where the total weight of a specific spin configuration is the product of the weights of the diagonal and vertical bonds as shown in Fig.~\ref{fig:mapping_main} (b). The weight of the diagonal bond is given by the inner product between two diagonally adjacent permutation spins
\begin{eqnarray}
    w_{d}(\sigma, \tau) = \langle \sigma \vert \tau \rangle = d^{r-\vert \sigma^{-1} \tau \vert},
\end{eqnarray}
and the weight of the vertical bond is given by the Weingarten function. Due to that $\text{Moeb}(\sigma)$ (Moebius number of $\sigma$) can be negative~\cite{collinsIntegrationRespectHaar2006}, we need to integrate out the $\tau$ spins to obtain positive three-body weights of downward triangles as shown in Fig.~\ref{fig:mapping_main} (d)
\begin{eqnarray}
    && W^{0}(\sigma_{1}, \sigma_{2}; \sigma_{3})  \\ \nonumber
    &=& \sum_{\tau \in S_{r}} \text{Wg}_{d^{2}}^{(r)} (\sigma_{3}\tau^{-1}) d^{2r-\vert \sigma_{1}^{-1} \tau \vert - \vert \sigma_{2}^{-1} \tau \vert}.
    \label{eq:3body}
\end{eqnarray}
Therefore, the total weight of a specific spin configuration is the product of the weights of the downward triangles as illustrated in Fig.~\ref{fig:mapping_main} (c). 

In the following discussion, we focus on the limit of large local Hilbert space dimension, $d \to \infty$, i.e., the partition function $Z$ is determined by the weight of the dominant spin configuration, which makes the analytical analysis easy and provides consistent theoretical understanding with numerical results with finite $d$ numerically validated below.

Before the discussion of the relation between the von Neumann entropy and free energy, we show the weights of the downward triangles with two specific spin configurations:
\begin{itemize}
    \item $\sigma_{1} = \sigma_{2} = \sigma_{3} = \sigma$: 
    \begin{eqnarray}
    W^{0}(\sigma,\sigma; \sigma) &=& \sum_{\tau \in S_{r}} \text{Wg}_{d^{2}}^{(r)}(\sigma \tau^{-1}) d^{2r-2\vert \sigma^{-1} \tau \vert} \\ \nonumber
    &\approx& \sum_{\tau \in S_{r}} \text{Moeb}(\sigma \tau^{-1}) d^{-4 \vert \sigma^{-1} \tau \vert} \\ \nonumber
    &\approx& d^{0}.
    \end{eqnarray}
    \item $\sigma_{1} = \sigma^{\prime}, \sigma_{2}=\sigma_{3}=\sigma$ or $\sigma_{2} = \sigma^{\prime}, \sigma_{1}=\sigma_{3}=\sigma$:
    \begin{eqnarray}
        \label{eq:domainwall}
        W^{0}(\sigma^{\prime},\sigma; \sigma) &=& W^{0}(\sigma,\sigma^{\prime}; \sigma) \\ \nonumber
        &=& \sum_{\tau \in S_{r}} \text{Wg}_{d^{2}}^{(r)}(\sigma \tau^{-1}) d^{2r-\vert \sigma^{-1} \tau \vert-\vert (\sigma^{\prime})^{-1} \tau \vert} \\ \nonumber
        &\approx& \sum_{\tau \in S_{r}} \text{Moeb}(\sigma \tau^{-1}) d^{-3 \vert \sigma^{-1} \tau \vert - \vert (\sigma^{\prime})^{-1} \tau \vert} \\ \nonumber
        &\approx& d^{-\vert (\sigma^{\prime})^{-1}) \sigma \vert}.
    \end{eqnarray}
\end{itemize}
While there are other possible spin configurations, the configuration that maximizes the triangle weight occurs when $\sigma_{1} = \sigma_{2} = \sigma_{3} = \sigma$. Therefore, the spin-spin interaction of the effective statistical model is ferromagnetic, and thus all the spins tend to be in the same direction to achieve the largest total weight. However, as discussed below, due to the particular top boundary conditions and the presence of quantum noises, the $S_{r}$ rotational symmetry is broken~\cite{BAO2021168618, Noise_bulk, jian2021quantum, PhysRevB.108.104310, PhysRevB.107.014307, PhysRevB.108.104310}
and domain walls may appear with unit energy of ${\rm log} (W^{0}(\sigma^{\prime},\sigma; \sigma))$ in the dominant spin configuration.

\subsection{Relation between the von Neumann entropy and the free energy}
Having established the mapping between the quantum circuit and the effective statistical model, we then introduce how to obtain the von Neumann entropy and mutual information of the circuit model from the free energy of the statistical model.

We first rewrite the von Neumann entropy $S_{\beta}$ of the subsystem $\beta$ ($\bar{\beta}$ represents the complementary region to $\beta$) as
\begin{eqnarray}
    S_{\beta} = \underset{n \rightarrow 1}{\lim} S^{(n)}_{\beta} = \underset{n \rightarrow 1}{\lim} \frac{1}{1-n} \mathbb{E}_{\mathcal{U}} \log\frac{\tr \rho_{\beta}^{n}}{(\tr \rho)^{n}},
\end{eqnarray}
where $\rho_{\beta}$ is the reduced density matrix of subsystem $\beta$ and $S^{(n)}_{\beta}$ is the $n$-th order Renyi entropy. In $n$-fold replicated Hilbert space, 
\begin{eqnarray}
    S_{\beta}^{(n)} &=& \frac{1}{1-n} \mathbb{E}_{\mathcal{U}} \log \frac{\tr \rho_{\beta}^{n}}{(\tr \rho)^{n}}  \\ \nonumber
    &=& \frac{1}{1-n} \mathbb{E}_{\mathcal{U}} \log \frac{\Tr((C_{\beta} \otimes I_{\bar{\beta}}) \rho^{\otimes n})}{\Tr ( (I_{\beta} \otimes I_{\bar{\beta}}) \rho^{\otimes n})}  \\ \nonumber
    &=& \frac{1}{1-n} \mathbb{E}_{\mathcal{U}} \log \frac{Z^{(n)}_{\beta}}{Z^{(n)}_{0}},
\end{eqnarray}
where $C_{\beta}$ and $I_{\beta}$ are the cyclic and the identity permutations among the $n$ ket indices of subsystem $\beta$ respectively, i.e.,
\begin{eqnarray}
    C_{\beta} &=& \otimes_{i \in \beta} C_{i}, \\ \nonumber
    I_{\beta} &=& \otimes_{i \in \beta} I_{i},
\end{eqnarray}
where $C_{i} = \begin{pmatrix} 1 & 2 & ... & n  \\ 2 & 3 & ... & 1\end{pmatrix}_{i} $ and $I_{i} = \begin{pmatrix} 1 & 2 & ... & n  \\ 1 & 2 & ... & n\end{pmatrix}_{i} $ are the cyclic and identity permutations among the $n$ ket indices of site $i$. Via the replica trick~\cite{nishimori2001statistical, kardar2007statistical}, we can overcome the difficulty of the average outside the logarithmic function
\begin{eqnarray}
    \mathbb{E}_{\mathcal{U}} \log Z_{\beta}^{(n)} &=& \underset{k \rightarrow 0}{\lim} \frac{1}{k} \log \{ \mathbb{E}_{\mathcal{U}} (Z_{\beta}^{(n)})^{k} \}  \\ \nonumber
    &=& \underset{k \rightarrow 0}{\lim} \frac{1}{k} \log Z_{\beta}^{(n,k)},\\ \nonumber 
    \mathbb{E}_{\mathcal{U}} \log Z_{0}^{(n)} &=& \underset{k \rightarrow 0}{\lim} \frac{1}{k} \log \{ \mathbb{E}_{\mathcal{U}} (Z_{0}^{(n)})^{k} \} \\ \nonumber
    &=& \underset{k \rightarrow 0 }{\lim} \frac{1}{k} \log Z_{0}^{(n,k)},
\end{eqnarray}
where 
\begin{eqnarray}
    \label{eq:topboundary}
    Z_{\beta}^{(n,k)} &=&\Tr \left\{ ( C_{\beta} \otimes I_{\bar{\beta}}) ^{\otimes k} ( \mathbb{E}_{\mathcal{U}} \rho^{\otimes nk} ) \right\}  \\ \nonumber
    &=& \Tr  \left\{ \mathbb{C}_{\beta} \otimes \mathbb{I}_{\bar{\beta}}  ( \mathbb{E}_{\mathcal{U}} \rho^{\otimes nk} ) \right\}, \\ \nonumber
    Z_{0}^{(n,k)} &=&\Tr \left\{ ( I_{\beta} \otimes I_{\bar{\beta}}) ^{\otimes k} ( \mathbb{E}_{\mathcal{U}} \rho^{\otimes nk} ) \right\} \\ \nonumber
    &=& \Tr \left\{ \mathbb{I}_{\beta} \otimes \mathbb{I}_{\bar{\beta}}  ( \mathbb{E}_{\mathcal{U}} \rho^{\otimes nk} ) \right\},
\end{eqnarray}
with $\mathbb{C} = C^{\otimes k} $ and $\mathbb{I} = I^{\otimes k} $ are permutations in the $r$-fold replicated Hilbert space with $r=nk$. Therefore,
\begin{eqnarray}
    S_{\beta} = \underset{n \rightarrow 1}{\underset{k \rightarrow 0 }{\lim}} \frac{1}{k(1-n)} \log \left\{ \frac{Z_{\beta}^{(n,k)}}{Z_{0}^{(n,k)}} \right\},
    \label{eq:smzz}
\end{eqnarray}
where $Z$ is the partition function of the classical spin model, corresponding to the weight of the dominant spin configuration with the largest weight of the ferromagnetic spin model in the large $d$ limit. Therefore, $S_{\beta}$ can be represented as the free energy difference:
\begin{eqnarray}
    S_{\beta}^{(n,k)} = \frac{1}{k(n-1)} \left[ F_{\beta}^{(n,k)} - F_{0}^{(n,k)}\right].
    \label{eq:fnk}
\end{eqnarray}
We note that the free energy $F^{(n,k)}$ is proportional to the length of the domain wall with unit energy $k(n-1)$, and thus $\frac{1}{k(n-1)} F^{(n,k)}$ is independent of the index $(n,k)$. Consequently, the limit to extract von Neumann entropy shown in Eq.~\eqref{eq:smzz} can be safely taken. Moreover, as illustrated in Eq.~\eqref{eq:topboundary}, the top boundary conditions are fixed to:
$\mathbb{C}_{\beta} \otimes \mathbb{I}_{\bar{\beta}}$ for $Z_{\beta}$ and $\mathbb{I}_{\beta} \otimes \mathbb{I}_{\bar{\beta}}$ for $Z_{0}$, and the bottom boundary condition is free with the initial product state. As a result, the dominant spin configuration contributing to $F_{0}^{(n,k)}$ is always that all the spins are fixed to $\mathbb{I}$ and $F^{(n,k)}_{0}$ is zero. Therefore, $S_{\beta}$ is determined by the free energy $F_{\beta}^{(n,k)}$.

In the absence of quantum noises, the dominant spin configuration contributing to $F^{(n,k)}_{AB}$ is that all the spins are fixed to $\mathbb{C}$ and the free energy $F_{AB}^{(n,k)}$ is zero. Therefore, $S_{AB}$ is zero consistent with the fact of a pure state.

\subsection{Noise-induced entanglement phase transitions}
In this section, we introduce the effects of quantum noises and the theoretical understanding of noise-induced entanglement phase transitions. We showcase by using the reset channel $\mathcal{R}$ to model quantum noise and the conclusion does not depend on the choice of the quantum channel.

In terms of the effective statistical model, the presence of a reset channel changes the weight of the diagonal bond between two diagonally adjacent spins,
\begin{eqnarray}
    \langle \sigma \vert \mathcal{R} \vert \tau \rangle = d^{r-\vert \tau \vert},
\end{eqnarray}
and thereby affects the weight of the downward triangle. The weight of the triangle with the same three spins is
    \begin{eqnarray}
        && W^{\mathcal{R}}(\sigma, \sigma; \sigma) \\ \nonumber
        &=& \sum_{\tau \in S_{r}} \text{Wg}^{(r)}_{d^{2}} (\sigma^{-1}\tau) d^{r-\vert \sigma^{-1} \tau \vert } \langle \sigma \vert \mathcal{R} \vert \tau \rangle \\ \nonumber
        &=& \sum_{\tau \in S_{r}} \text{Wg}^{(r)}_{d^{2}} (\sigma^{-1}\tau) d^{r-\vert \sigma^{-1} \tau \vert } d^{r-\vert \tau \vert} \\ \nonumber
        &\sim& d^{-\vert \sigma \vert}.
    \end{eqnarray}
Consequently, in the classical spin model, the quantum noise acts as magnetic field pinning in the direction $\mathbb{I}$ and the random space-time locations of quantum noises can be treated as quenched disorders. 

For the classical spin model corresponding to $S_{AB}$, the dominant spin configuration for $\alpha>1$ is that all spins are fixed to $\mathbb{C}$ as shown in Fig. \ref{fig:statistical} (a), same as the noiseless case and the free energy is proportional to $qLT$, the average number of quantum noises, due to the energy cost arising from the magnetic field. However, when $\alpha<1$, this configuration is not favored compared to the spin configuration with a domain wall as shown in Fig. \ref{fig:statistical} (b), whose free energy is proportional to the domain wall length as $s_{0}L$ where $s_{0}$ is a constant. The domain wall is formed as follows~\cite{PhysRevLett.132.240402}. Due to the fixed top boundary condition of the classical spin model corresponding to $S_{AB}$, the classical spins remain $\mathbb{C}$ until the reversed evolution from the top to the bottom encounters a quantum noise $N(x_{1},t_{1})$. Spins inside the downward light cone of $N(x_{1},t_{1})$ will change from $\mathbb{C}$ to $\mathbb{I}$ while other spins are unchanged. 
Other quantum noises inside the light cone of $N(x_{1}, t_{1})$ do not affect the spin configuration as these spins are already in the $\mathbb{I}$ domain, while another quantum noise outside the light cone, e.g., $N(x_{2},t_{2})$, will also change the spins within its respective backward light cone from $\mathbb{C}$ to $\mathbb{I}$. As a result, the domain wall is composed of the downward light cone of topmost noises, i.e., the inflection points of the domain wall correspond to the locations of topmost noises as shown in Fig.~\ref{fig:statistical} (b) and other noises remain in domain $\mathbb{I}$ with no extra energy contribution. Moreover, there is an effective length scale $L_{\text{eff}} \sim q^{-1}$ determined by the average distance between adjacent quantum noises, as shown in Fig.~\ref{fig:statistical} (b). 

Besides, the spin configuration with a domain wall is always favored for the classical spin model corresponding to $S_{A(B)}$ due to the fixed top boundary conditions. 
Consequently, when $\alpha=1$, a competition between the two types of spin configurations of $S_{AB}$ arises resulting in a volume law entanglement phase with the mutual information proportional to $(s_{0}-p \frac{T}{L})L$ when $p<s_{0}L/T$ and an area law entanglement phase when $p>s_{0}L/T$, corresponding to the vertical line at $p_{m}=0$ shown in Fig.~\ref{fig:setup} (c).

The projective measurements will not alter the mechanism of noise-induced entanglement phase transition, as given by the competition between two candidate spin configurations of $S_{AB}$. However, the projective measurements, regarded as attractive random Gaussian potential in the effective statistical model, can render the domain wall fluctuating to go through more measurements to minimize the free energy. Consequently, building on our previous works~\cite{PhysRevB.107.L201113, PhysRevLett.132.240402}, the free energy of the domain wall can be obtained from the KPZ theory including a subleading term proportional to $L_{\text{eff}}^{1/3}$. Therefore, when the noise probability $p>p_c$, the entanglement obeys a power law scaling in the presence of projective measurements, as shown in the blue region in Fig.~\ref{fig:setup} (c).

Notably, this entanglement phase transition is a first-order phase transition arising from the competition between two spin configurations. For $\alpha>1$ ($\alpha<1$) the spin configuration with all spins fixed to $\mathbb{C}$ (with domain wall) always dominates, leading to a single volume law (power-law or area-law) entanglement phase. 
Therefore, this noise-induced phase transition only occurs with $\alpha=1$.

\subsection{Noise-induced coding transition}

\begin{figure}[ht]
\centering
\includegraphics[width=0.46\textwidth, keepaspectratio]{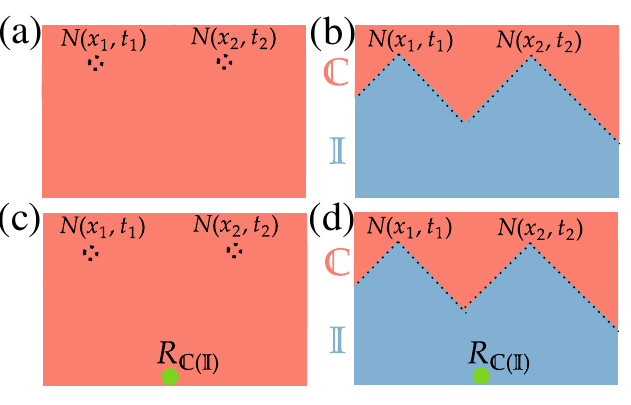}
\caption{The spin configurations of the effective statistical model: red for $\mathbb{C}$ and blue for $\mathbb{I}$. (a)(b) show the two competing spin configurations for $S_{AB}$ in the entanglement phase transition. The effective length scale of the domain wall shown in (b) is determined by the average distance between adjacent quantum noises. In the presence of projective measurements, the domain wall will fluctuate away from its original path. (c)(d) show the two competing spin configurations in the coding transition. The spin corresponding to the Bell pair is denoted as $R$ and is fixed to $\mathbb{I}$ and $\mathbb{C}$ for $S_{AB}$ and $S_{AB \cup R}$ respectively. $N$ represents the topmost quantum noise and other quantum noises in the bulk are not shown here.}
\label{fig:statistical}
\end{figure}

Next, we apply the statistical model understanding to the noise-induced coding transition. In addition to the particular top boundary conditions discussed above, the spin at the bottom with the Bell pair is fixed by the top boundary conditions of the reference qubit: $\mathbb{I}$, $\mathbb{C}$, $\mathbb{C}$ for $S_{AB}$, $S_{R}$ and $S_{AB \cup R}$, respectively, see Fig. \ref{fig:statistical} (c) and (d). 
$S_{R}$ remains constant because the dominant spin configuration is $\mathbb{I}$ in the bulk, regardless of the noise probability. The defect created at the bottom due to the Bell pair contributes to the free energy $1$. 
However, for $S_{AB}$ and $S_{AB \cup R}$, the competition between the spin configuration with all spins fixed to $\mathbb{C}$ and the spin configuration with a topmost domain wall, as shown in Fig. \ref{fig:statistical} (c) and (d) still exists. The mutual information $I_{AB \cup R} =2$ when the former spin configuration dominates while $I_{AB \cup R} =0$ when the latter dominates. As a result, the encoded information is perfectly protected with the noise probability prefactor below the critical value and a noise-induced coding transition occurs as the noise probability prefactor $p$ increases. Via the understanding of the unified statistical model, the coding transition is demonstrated to share the same critical value and exponent as the entanglement phase transition. 

We note that the coding transition induced by bulk noises differs significantly from the spatial boundary noises case where the information is only partially protected below the critical value~\cite{Coding_Vijay}. In Sec.~\ref{sec:distinction}, we present a detailed discussion of the distinction in the coding transition between bulk noises and boundary noises.

\section{Numerical results}
\label{sec:numerical}
To support our theoretical understanding, we conduct extensive simulations of large-scale stabilizer circuits where random Clifford two-qubit unitary gates form a unitary $3$-design~\cite{Unitary3design, bergSimpleMethodSampling2021} and thus give qualitatively similar entanglement behaviors as the Haar random gates. 
To model the quantum noise, we employed the reset channel defined as follows 
\begin{eqnarray}
    \mathcal{R}_{i}(\rho) = \tr_{i}(\rho) \otimes \vert 0 \rangle \langle 0 \vert_{i}.
\end{eqnarray}
We note that the conclusions are independent of the choice of the quantum channels~\cite{PhysRevLett.132.240402}.

\begin{figure}[ht]
\centering
\includegraphics[width=0.49\textwidth, keepaspectratio]{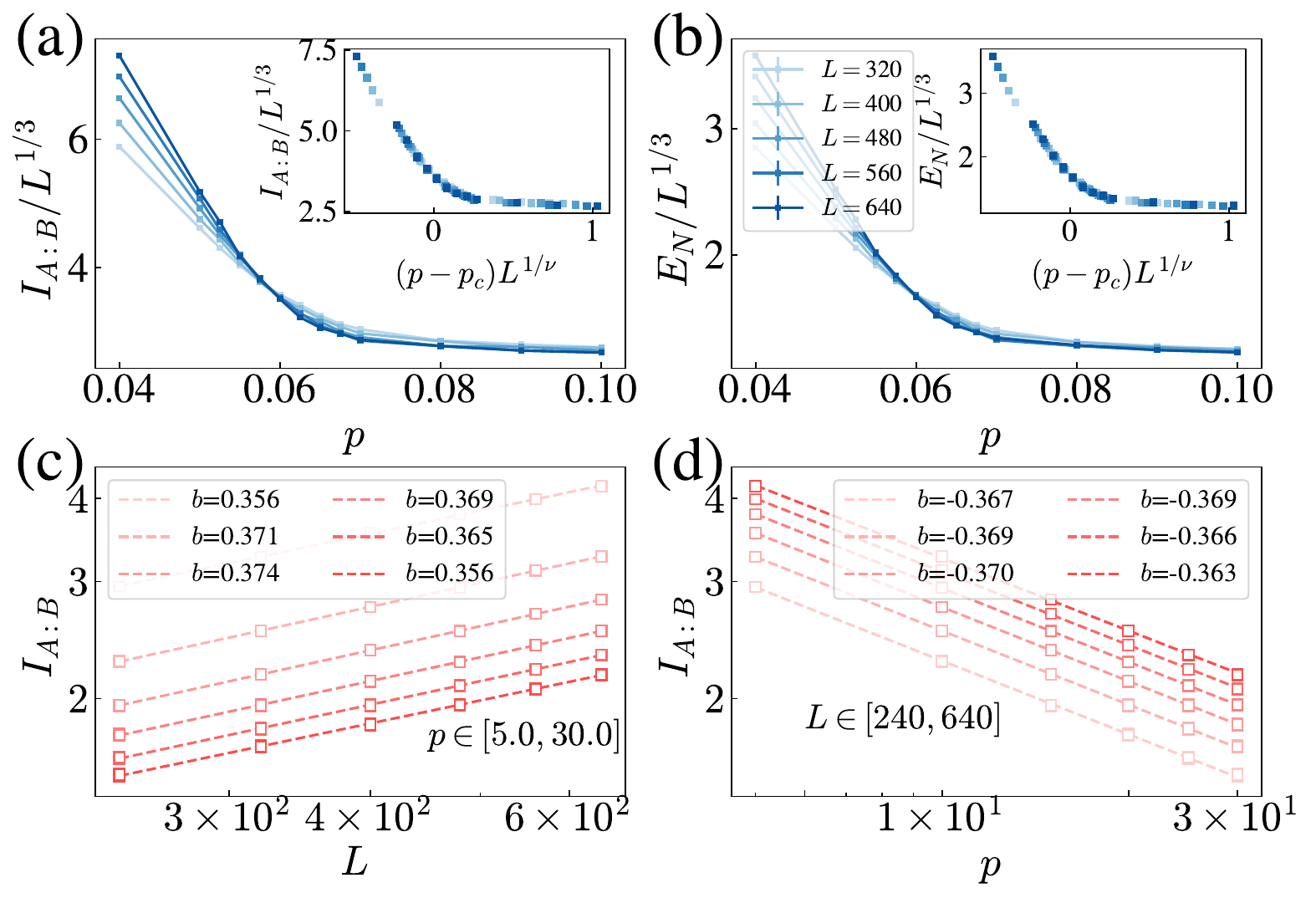}
\caption{Here, $q=p/L$, $p_{m}=0.2$ and $T=4L$. (a) shows the rescaled mutual information $I_{A:B}/L^{1/3}$ vs noise probability prefactor $p$; (b) shows the rescaled logarithmic entanglement negativity $E_{N}/L^{1/3}$ vs noise probability prefactor $p$. There is a noise-induced entanglement phase transition from a volume law phase to a power law phase with an increase of $p$. The insets show the data collapse with $p_{c}=0.0593$ and $\nu=2$. (c)(d) show the fitting of the mutual information. The obtained power is very close to the theoretical predictions, showing that $I_{A:B} \sim (L/p)^{1/3}$.}
\label{fig:reset_pm0.2_ET}
\end{figure}

We set $\alpha=1$ and $p_{m}<p_{m}^{c}$. 
The numerical results of $E_{N}$ and $I_{A:B}$ with $p_{m}=0.2$ and varying noise probabilities are shown in Fig. \ref{fig:reset_pm0.2_ET} (a) and (b). 
The $y$-axis represents the rescaled entanglement, denoted as $E_{N}/L^{1/3}$ or $I_{A:B}/L^{1/3}$. 
In the power-law entanglement phase with large noise probabilities, the data obtained from different system sizes should collapse onto the same curve. 
Conversely, in the volume entanglement phase, the rescaled entanglement should increase as the system size increases. We observe a crossing point at a critical probability, $p_{c}$, indicating the noise-induced entanglement phase transition. 
To determine this critical probability, we employ data collapse with a scaling function 
\begin{eqnarray}
    S(p,L)/L^{1/3} = F\left((p-p_{c})L^{1/\nu} \right), 
\end{eqnarray}
where $S$ represents $E_{N}$ or $I_{A:B}$, $\nu$ is the critical exponent fixed to $2$ arising from the randomness of quantum noises~\cite{Coding_Vijay}. The data collapse is shown in the insets of Fig.~\ref{fig:reset_pm0.2_ET}. 
Similarly, we show $I_{A:B} \sim (L/p)^{1/3}$ scaling as illustrated in Fig. \ref{fig:reset_pm0.2_ET} (c) and (d). 
Therefore, we have demonstrated the noise-induced entanglement phase transition from a volume-law phase to a power-law phase. In the absence of projective measurements, there is an area-law entanglement phase instead of a power-law entanglement phase when the noise probability $p>p_c$, see more numerical results of the noise-induced entanglement phase transition in Appendix.~\ref{sec:numerical_nipt}.

\begin{figure}[t]
\centering
\includegraphics[width=0.46\textwidth, keepaspectratio]{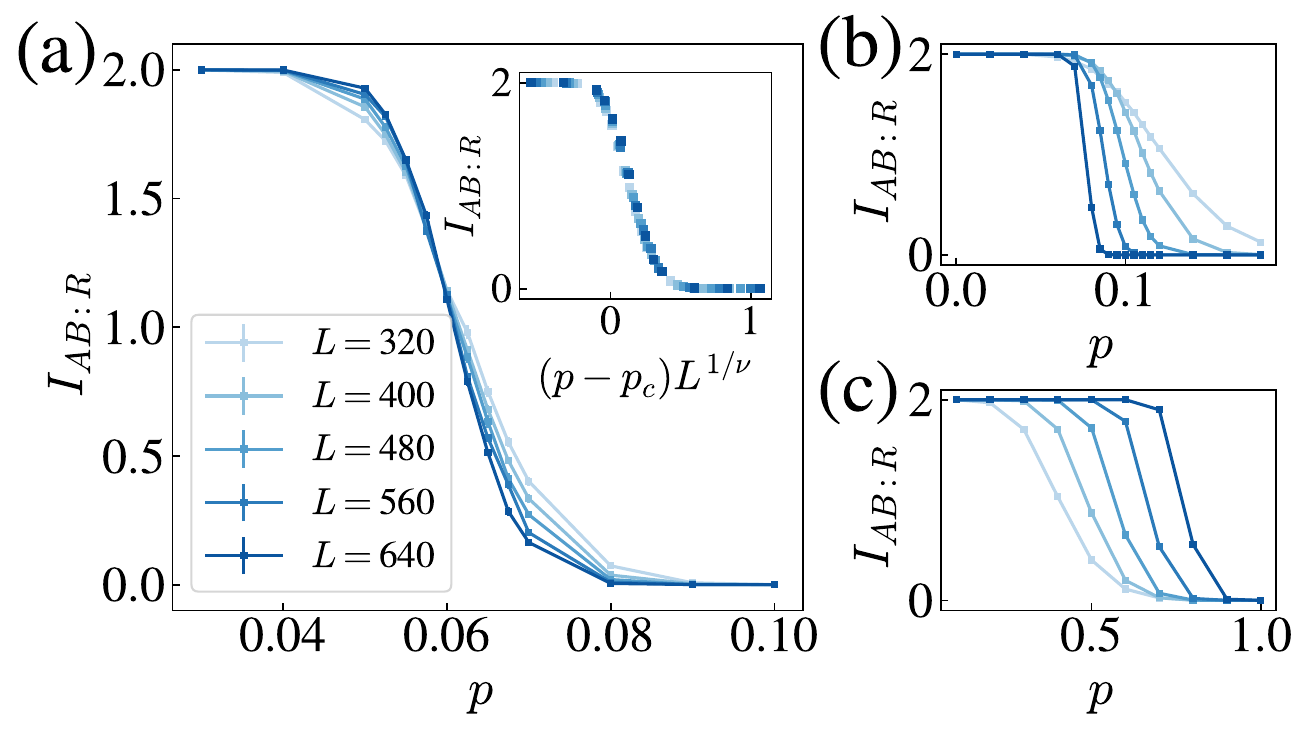}
\caption{(a) shows the noise-induced coding transition in the presence of quantum noises with $\alpha=1$. $p_{m}=0.2$ and $T=4L$. The inset shows the data collapse with critical probability $p_{c}=0.0543(73)$ and critical exponent $\nu=2.052(556)$ which are consistent with those of noise-induced entanglement phase transition. 
(b)(c) show the mutual information $I_{AB:R}$ with $\alpha=0.8$ and $\alpha=1.2$ respectively. The noise-induced phase transitions disappear when $\alpha \neq 1$.}
\label{fig:reset_pm0.2_Coding}
\end{figure}

Next, we numerically investigate the noise-induced coding transition. The numerical results of $I_{AB:R}$ are shown in Fig. \ref{fig:reset_pm0.2_Coding} (a). 
When the quantum noises are sparse with a small probability, the encoded quantum information can be perfectly protected in the thermodynamic limit, i.e., $I_{AB:R}=2$, consistent with the theoretical prediction. When $p$ is large, the information is destroyed, and $I_{AB:R}=0$. 
The inset of Fig. \ref{fig:reset_pm0.2_Coding} (a) shows the data collapse, where the obtained critical probability and exponent are consistent with those obtained from the noise-induced entanglement phase transition. See more numerical results of the noise-induced coding transition in Appendix.~\ref{sec:numerical_coding}.

Apart from the initial state encoding scheme, we also investigate the information protection from the steady-state encoding scheme. In this case, the information protection time scale is predicted to be $(L/p)^{1/2}$~\cite{PhysRevLett.132.240402}, which is consistent with our numerical results as shown in Appendix.~\ref{sec:numerical_nipt}.

Furthermore, as discussed above, the scaling exponent $\alpha=1$ is crucial for these noise-induced phase transitions. 
Here, we show the numerical results of mutual information $I_{AB:R}$ with $\alpha=0.8$ and $\alpha=1.2$ in Fig. \ref{fig:reset_pm0.2_Coding} (b) and (c), respectively. 
The encoded information will always be destroyed (perfectly protected) in the presence of quantum noises when $\alpha<1$ ($\alpha>1$) in the thermodynamic limit. 

\section{Distinction between bulk noises and boundary noises}
\label{sec:distinction}

In this section, we clarify the differences between the setup investigated in this work with bulk quantum noises and the setup investigated in Ref. \cite{Coding_Vijay} with spatial boundary quantum noises. The different space-time distributions of quantum noises result in qualitative differences in information protection and phase diagram of the coding transition.

\begin{figure}[ht]
\centering
\includegraphics[width=0.48\textwidth, keepaspectratio]{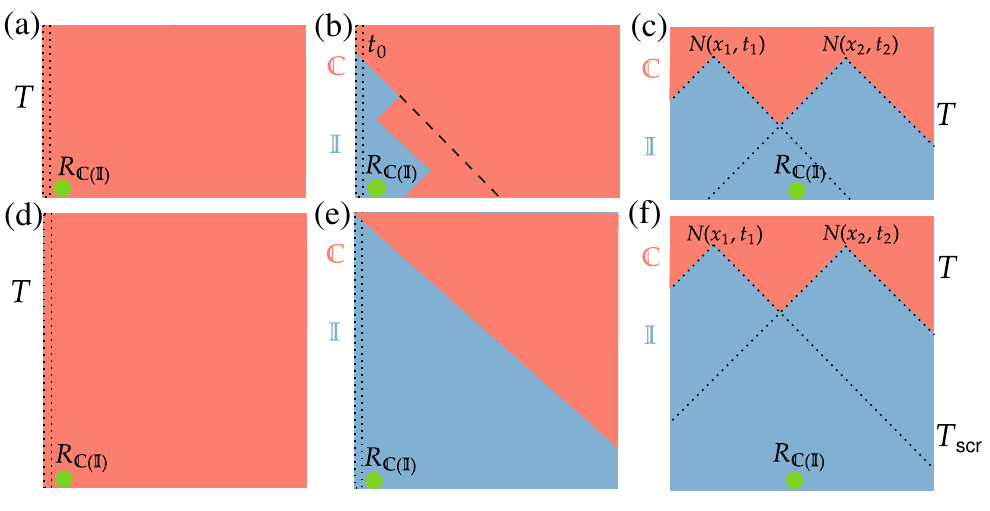}
\caption{(a) and (d) show the spin configurations with all spins fixed to $\mathbb{C}$ with $T/L<1$ and $T/L>1$ respectively. The quantum noises are at the left spatial boundary (dashed rectangle). (b) and (e) show the spin configurations with a domain wall. In (b), the black line shows another possible path for the domain wall with the same $t_{0}$. (c) and (f) show the domain wall configuration with quantum noises in the bulk without and with the scrambling process respectively.}
\label{fig:boundary_and_bulk}
\end{figure}

For the setup with quantum noises on the left spatial boundary~\cite{Coding_Vijay}, the location of encoded information is crucial and is set to nearest to the left boundary. We note that the information dynamics is agnostic with the location of encoded information for the bulk noises case even with open boundary conditions. Firstly, we analyze the competition between different candidate spin configurations in the large $d$ limit to provide analytical predictions of the coding transition with quantum noises on the left boundary. Since $S_{R}$ remains constant, we only consider the classical spin models corresponding to $S_{AB}$ and $S_{AB \cup R}$. One candidate spin configuration is that all spins are fixed to $\mathbb{C}$ (see Fig. \ref{fig:boundary_and_bulk} (a) and (d)) resulting in a free energy (on average) of $(qT+1) \vert \mathbb{C} \vert$ and $qT \vert \mathbb{C} \vert$ for $S_{AB}$ and $S_{AB \cup R}$ respectively. Consequently, the mutual information between the system and the reference qudit is $I_{AB:R} = 2 \log(d)$, indicating that the encoded information is perfectly protected. 
On the other hand, when $T/L<1$, another candidate spin configuration is to create a domain wall that starts from the left boundary at time $t_{0}$ and is annihilated at the bottom such that the Bell pair lives in the domain $\mathbb{I}$ (see Fig.~\ref{fig:boundary_and_bulk} (b)), while when $T/L >1$, another candidate spin configuration is to create a domain wall that starts from the left-top corner and is annihilated by the right boundary (see Fig.~\ref{fig:boundary_and_bulk} (e)). The mutual information in both cases is zero, i.e., the encoded information is destroyed by the quantum noises. However, the free energy of the former spin configuration (Fig.~\ref{fig:boundary_and_bulk} (b)) is $(qt_{0} + (T-t_{0})) \vert \mathbb{C} \vert$, which is larger than that with all spins fixed to $\mathbb{C}$ regardless of the choice of $t_{0}$. Consequently, the boundary noise induced coding transition is absent when $T/L<1$ in the large $d$ limit. On the contrary, the free energy of the latter spin configuration (Fig.~\ref{fig:boundary_and_bulk} (e)) is $L \vert \mathbb{C} \vert$ and thus the boundary noise induced coding transition from an information perfectly protected phase to an information lost phase occurs in the large $d$ limit as boundary noise probability increases.

However, as investigated in Ref.~\cite{Coding_Vijay}, the case with finite $d$ differs significantly from the theoretical predictions in the large $d$ limit. The boundary noise induced coding transition always exists regardless of the choice of $T/L$, although it is a first-order transition when $T/L>1$ while it is a second-order transition when $T/L<1$. Moreover, the information is partially protected with noise probability below the critical point, different from the theoretical prediction of perfect protection in the large $d$ limit. To understand this difference between large $d$ and finite $d$, we note that the path of the domain wall with fixed $t_{0}$ is not unique (see Fig. \ref{fig:boundary_and_bulk} (b)) and thus there is an entropy contribution to the free energy at finite temperature, i.e., finite $d$, which is crucial for the boundary noise induced coding transition with finite $d$. Moreover, as discussed in Ref.~\cite{Coding_Vijay}, the introduction of a pre-scrambling process before the noisy evolution will cause the disappearance of the second-order transition and the information will be perfectly protected when the noise probability is below the critical point, although it will not change the analysis above in the large $d$ limit.

For the setup considered in this work, when the scaling exponent of quantum noise is $\alpha=1$, i.e., $q=p/L$, the free energy (on average) of the spin configuration with all spins fixed to $\mathbb{C}$ is $(pT+1) \vert \mathbb{C} \vert$ and $pT \vert \mathbb{C} \vert$ for $S_{AB}$ and $S_{AB \cup R}$ respectively. However, the free energy of the spin configuration with a domain wall is always $O(L\vert \mathbb{C} \vert)$ regardless of the choice of $T/L$. Consequently, the noise-induced coding transition always exists in the large $d$ limit. Furthermore, the dominant domain wall configuration in our case is horizontal-like, contrasting to the vertical-like domain wall in the boundary noises case. Therefore, the domain wall in the bulk noises case is unique because of the unitary constraint, and the analytical results from the large $d$ limit match well with the numerical results from finite $d=2$, including the consistent critical point and perfect information protection. Consequently, for the case with the bulk quantum noises investigated in this work, the results remain qualitatively the same with and without the pre-scrambling stage. See more numerical results in Appendix.~\ref{sec:numerical_coding}.

\begin{figure}[ht]
\centering
\includegraphics[width=0.4\textwidth, keepaspectratio]{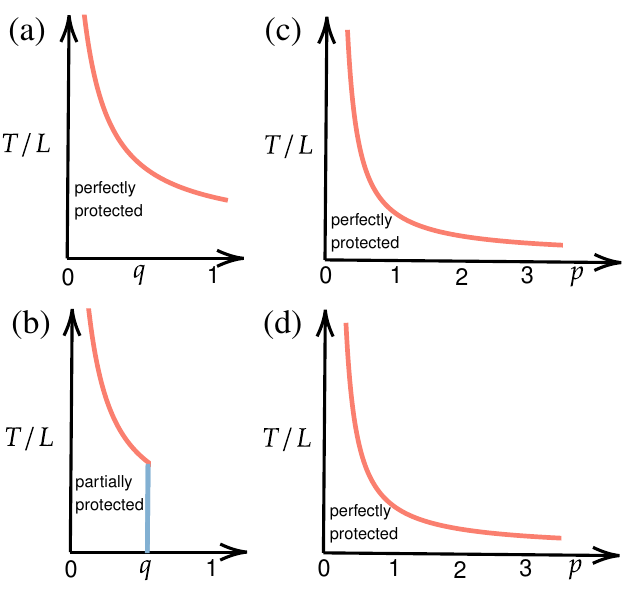}
\caption{Schematic phase diagrams for coding transition without pre-scrambling process. (a) left-boundary noises and $d=\infty$; (b) left-boundary noises and $d=2$~\cite{Coding_Vijay}. (c) bulk noises and $d=\infty$; (d) bulk noises and $d=2$. In the presence of bulk quantum noises, the schematic phase diagrams are the same for $d=2$ and $d=\infty$. The red line represents a first-order transition while the blue line represents a second-order phase transition.}
\label{fig:finite_large_d}
\end{figure}

In conclusion, the schematic phase diagrams with infinite $d$ and finite $d$ for these two setups
are shown in Fig. \ref{fig:finite_large_d}. 

\section{Conclusions and discussion}
\label{sec:conclusion}
We have investigated the noise-induced entanglement phase transition and coding transition in the presence of quantum noises with scaling exponent $\alpha=1$. 
Theoretical analysis reveals that these phase transitions can be understood as the first-order phase transition by the competition between different spin configurations within an effective statistical model. 
Through numerical simulations, we have validated these noise-induced phase transitions and their critical behaviors, generalizing the framework of MIPTs to the cases with quantum noises as shown in Fig. \ref{fig:setup}. 
Additionally, the influence of the scaling exponents $\alpha$ has also been discussed. We have also investigated the disappearance of phase transition when $\alpha \neq 1$ as only one spin configuration dominates regardless of $p$.

The power law scaling with different exponents in open quantum systems can lead to drastic changes of dynamical phases and phase transitions. In this Letter, we observe that different power law scalings for noise strength lead to volume law and area law phases, separated by the $\alpha=1$ regime with an entanglement phase transition. Similarly, different power law scalings for noise spectrum result in sub-Ohmic and super-Ohmic regimes, separated by the Ohmic regime with a delocalized-localized transition~\cite{RevModPhys.59.1, weiss2012quantum}. 
It is an interesting future direction to investigate the distinction and connection between general dissipation dynamics phenomena and the noise-induced transitions reported here.

Furthermore, we note that the noise-induced entanglement or coding transitions have a slightly different statistical model picture compared to the noise-induced computational complexity transition in random circuit sampling~\cite{NIPT1_arxiv, NIPT2_arxiv, PhysRevA.109.042414, PhysRevA.109.042414}. 
In the latter case, there are only two replicas, and the critical probability $p_{c}$ is independent of the choice of the ratio $L/T$ as analytically predicted as $p_c\approx 1$. 
However, for noise-induced entanglement or coding transitions discussed in this Letter, $p_{c} \sim L/T$. 
For the infinite time limit $L/T \rightarrow 0$, these noise-induced phase transitions vanish, consistent with the fact that the encoded information is ultimately destroyed in the presence of quantum noises. 
To demonstrate these differences, we have conducted simulations on noise-induced computational complexity transition in Clifford circuits as shown in Appendix.~\ref{sec:complexity}. 

\section{Acknowledgement}
This work is supported in part by the MOSTC under Grant No. 2021YFA1400100, the Innovation Program for Quantum Science and Technology (grant No. 2021ZD0302502), the NSFC under Grants No. 12347107 and No. 12334003. S.L. and M.R.L. acknowledge the support from the Lavin-Bernick Grant during the visit to Tulane University, where part of the work was conducted. The work of S.K.J. is supported by a startup fund at Tulane University. The work of S.X.Z. is supported in part by a startup grant at IOP-CAS. H.Y. acknowledges the support by the Xplorer Prize through the New Cornerstone Science Foundation.  

\appendix
\setcounter{equation}{0}
\renewcommand{\thefigure}{S\arabic{figure}}
\setcounter{figure}{0}

\section{Numerical results for noise-induced entanglement phase transition}
\label{sec:numerical_nipt}
\begin{figure}[ht]
\centering
\includegraphics[width=0.49\textwidth, keepaspectratio]{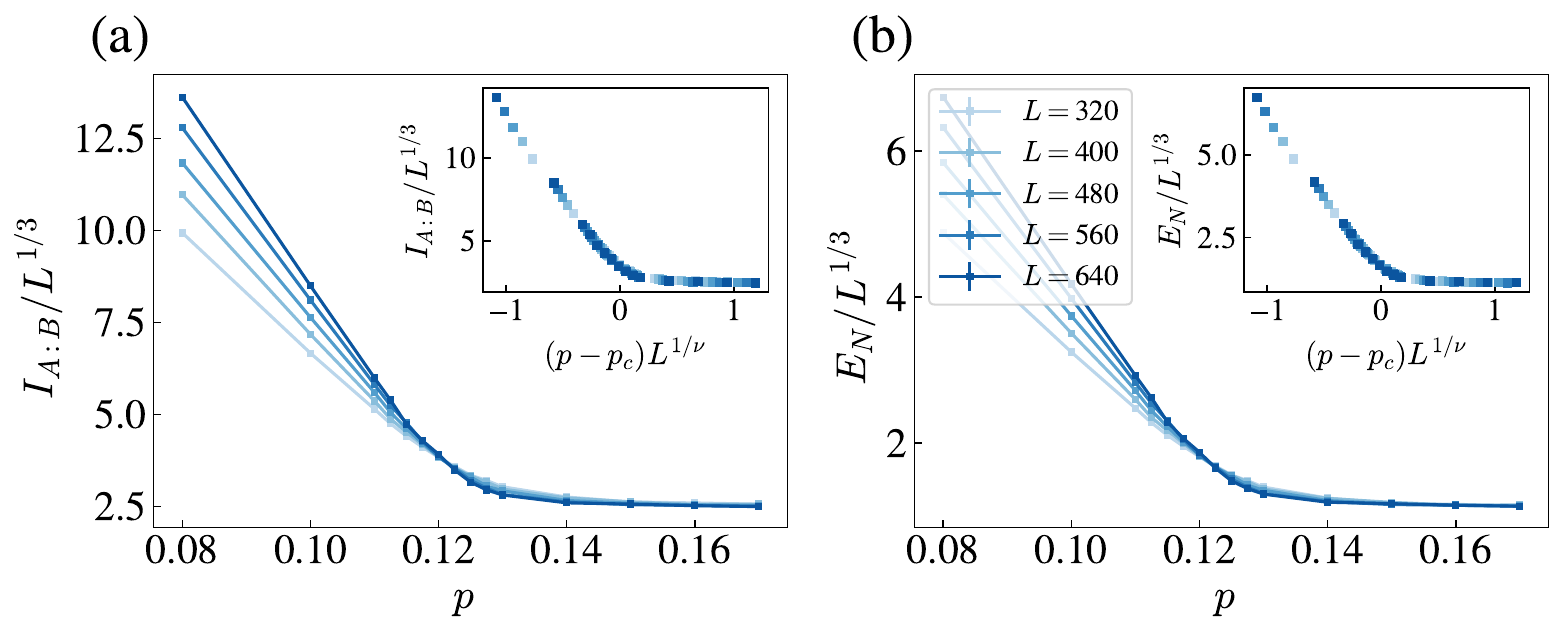}
\caption{The probability of reset channels is $q=p/L$ and the probability of measurements is $p_{m}=0.1$. We set $T=4L$. (a) shows the rescaled mutual information within the system $I_{A:B}/L^{1/3}$ vs noise probability $p$; (b) shows the rescaled logarithmic entanglement negativity within the system $E_{N}/L^{1/3}$ vs noise probability $p$. The insets show the data collapse with $p_{c}=0.123$ and $\nu=2$}
\label{fig:reset_pm0.1_ET}
\end{figure}

\begin{figure}[ht]
\centering
\includegraphics[width=0.49\textwidth, keepaspectratio]{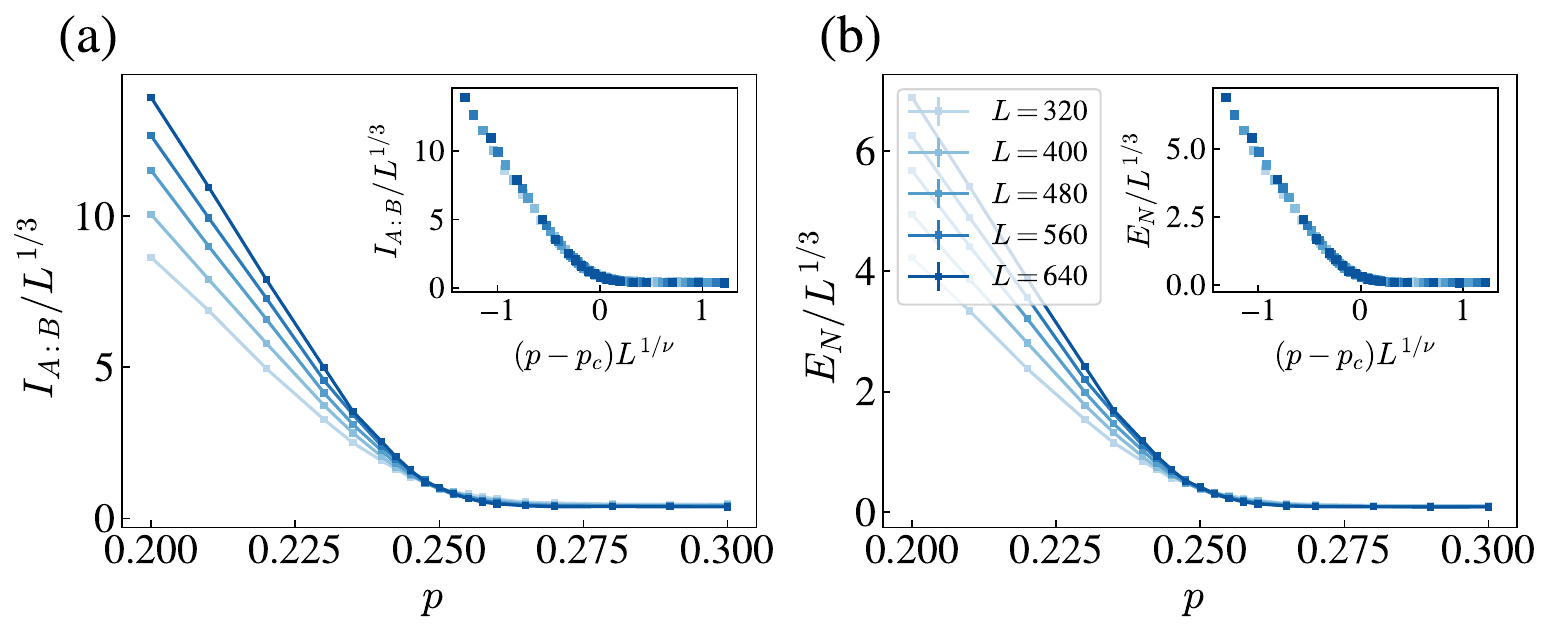}
\caption{The probability of reset channels is $q=p/L$ and the probability of measurements is $p_{m}=0.0$. We set $T=4L$. (a) shows the rescaled mutual information within the system $I_{A:B}/L^{1/3}$ vs noise probability $p$; (b) shows the rescaled logarithmic entanglement negativity within the system $E_{N}/L^{1/3}$ vs noise probability $p$. The insets show the data collapse with $p_{c}=0.252$ and $\nu=2$.}
\label{fig:reset_pm0.0_ET}
\end{figure}

\begin{figure}[ht]
\centering
\includegraphics[width=0.49\textwidth, keepaspectratio]{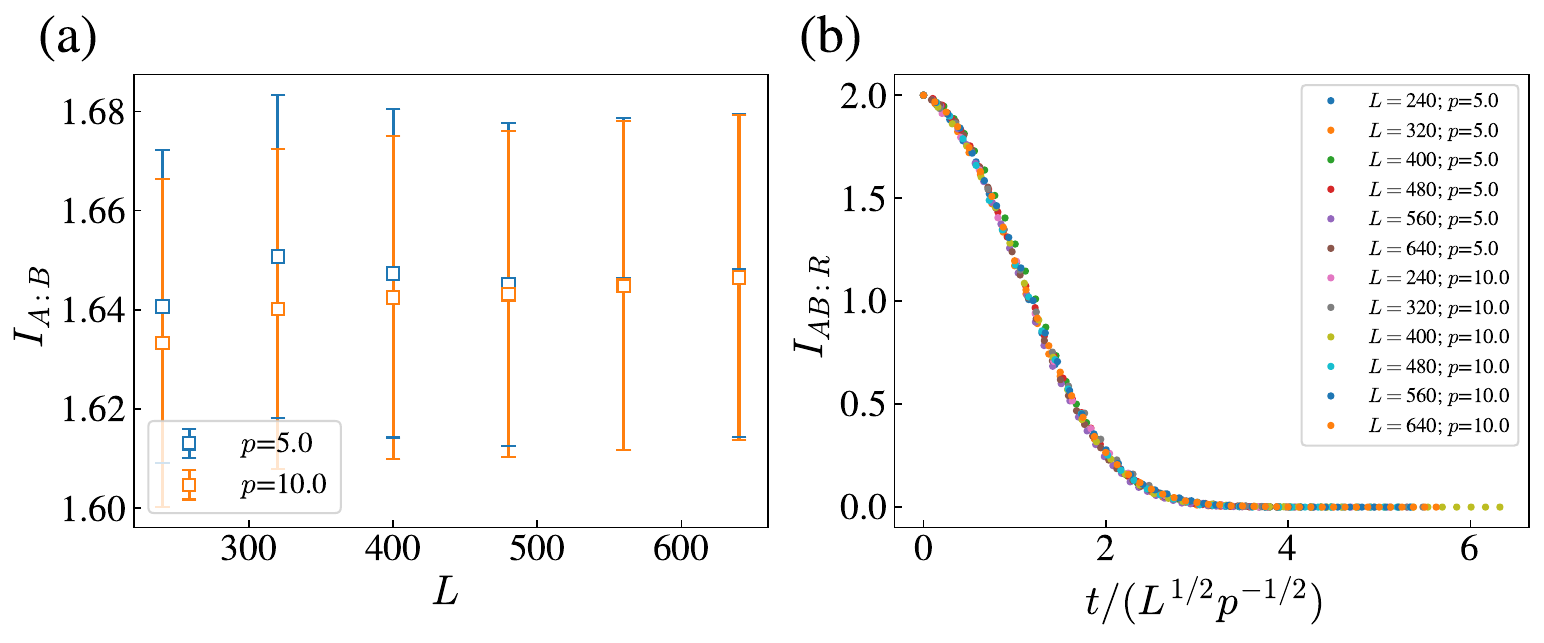}
\caption{The probability of reset channels is $q=p/L$ and the probability of measurements is $p_{m}=0.0$. (a) shows the mutual information $I_{A:B}$ of the steady states vs system size $L$ with the probability of noise $p>p_{c}$. The entanglement within the system obeys area law. (b) shows the dynamics of mutual information $I_{AB:R}$ vs rescaled time $t/(L/p)^{1/2}$.} 
\label{fig:reset_pm0.0_IAR}
\end{figure}

In this section, we present additional numerical results of the noise-induced entanglement phase transition. The numerical results for the mutual information and logarithmic entanglement negativity with fixed measurement probability $p_{m}=0.1$ and $T/L=4$ are shown in Fig. \ref{fig:reset_pm0.1_ET}. In the absence of projective measurement, i.e., $p_{m}=0.0$, the noise-induced entanglement phase transition still occurs. However, when the probability of noise $p$ exceeds the critical value $p_{c}$, the entanglement within the system follows an area law, as illustrated in Fig. \ref{fig:reset_pm0.0_ET} and Fig. \ref{fig:reset_pm0.0_IAR} (a).

\begin{figure}[ht]
\centering
\includegraphics[width=0.35\textwidth, keepaspectratio]{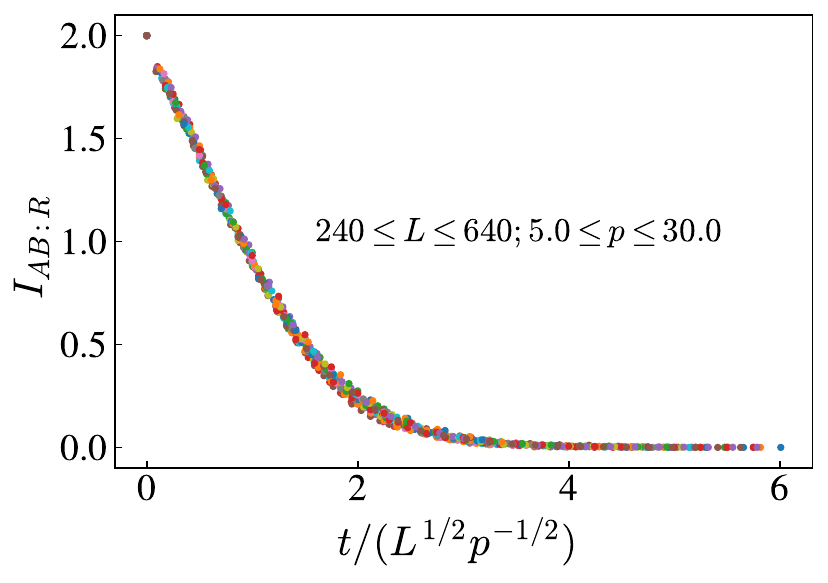}
\caption{The probability of reset channels is $q=p/L$ and the probability of measurements is $p_{m}=0.2$. The dynamics of mutual information $I_{AB:R}$ can be collapsed with rescaled time $t/(L/p)^{1/2}$.}
\label{fig:reset_pm0.2_IAR}
\end{figure}

Moreover, to validate our analytical understanding, we also investigate the timescale of information protection for the steady states in this noise-induced power or area law entanglement phase. As discussed in our previous work~\cite{PhysRevLett.132.240402}, this timescale is $q^{-1/2}$, i.e., $(L/p)^{1/2}$ when $p$ is much larger than $p_{c}$ and can be understood as the analogy of the Hayden-Preskill protocol for black holes~\cite{PatrickHayden_2007} in noisy hybrid quantum circuits. Consequently, the dynamics of mutual information $I_{AB:R}$ can be collapsed with rescaled time $t/(L/p)^{1/2}$. The numerical results with $p_{m}=0.2$ and $p_{m}=0.0$ are shown in Fig. \ref{fig:reset_pm0.2_IAR} and Fig. \ref{fig:reset_pm0.0_IAR} (b) respectively.

\section{Numerical results for noise-induced coding transition}
\label{sec:numerical_coding}
\begin{figure}[ht]
\centering
\includegraphics[width=0.35\textwidth, keepaspectratio]{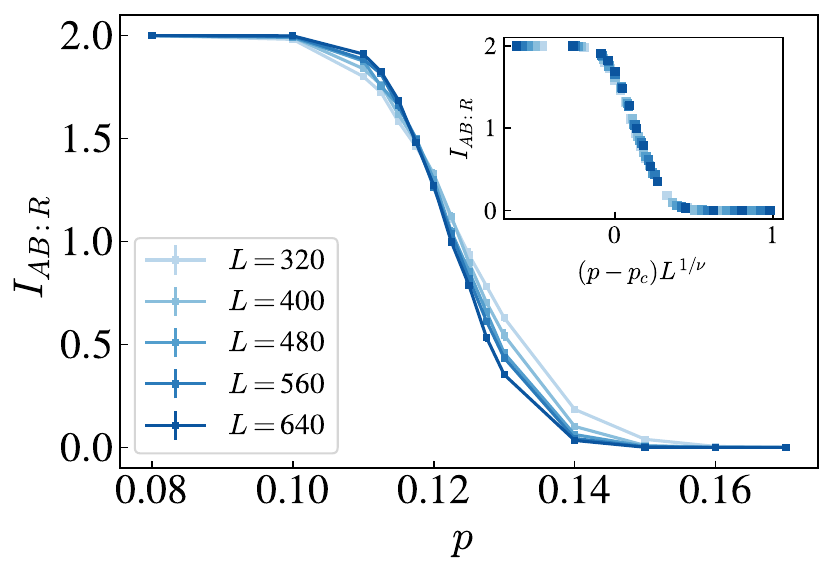}
\caption{The probability of reset channels is $q=p/L$ and the probability of measurements is $p_{m}=0.1$. We set $T_{\text{scr}}=T$ and $T=4L$. Inset shows the data collapse with $p_{c}=0.115(6)$ and $\nu=2.241(487)$.} 
\label{fig:reset_pm0.1_Coding}
\end{figure}

\begin{figure}[ht]
\centering
\includegraphics[width=0.35\textwidth, keepaspectratio]{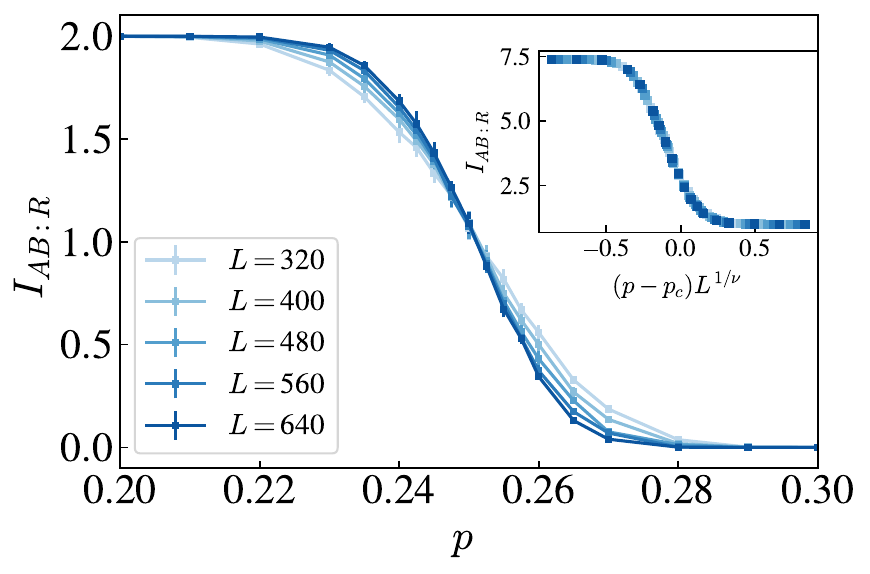}
\caption{The probability of reset channels is $q=p/L$ and the probability of measurements is $p_{m}=0.0$. We set $T_{\text{scr}}=T$ and $T=4L$. Inset shows the data collapse with $p_{c}=0.251(2)$ and $\nu=2.280(356)$.}
\label{fig:reset_pm0.0_Coding}
\end{figure}

In this section, we present additional numerical results of the noise-induced coding transition. The numerical results of mutual information $I_{AB:R}$ between the system ($AB$) and the reference qubit $R$ with measurement probabilities $p_{m}=0.1$ and $p_{m}=0.0$ are shown in Fig. \ref{fig:reset_pm0.1_Coding} and Fig. \ref{fig:reset_pm0.0_Coding} respectively. The critical exponent $\nu$ is close to $2$ and the critical probability $p_{c}$ increases as the probability of measurements decreases.

\begin{figure}[ht]
\centering
\includegraphics[width=0.49\textwidth, keepaspectratio]{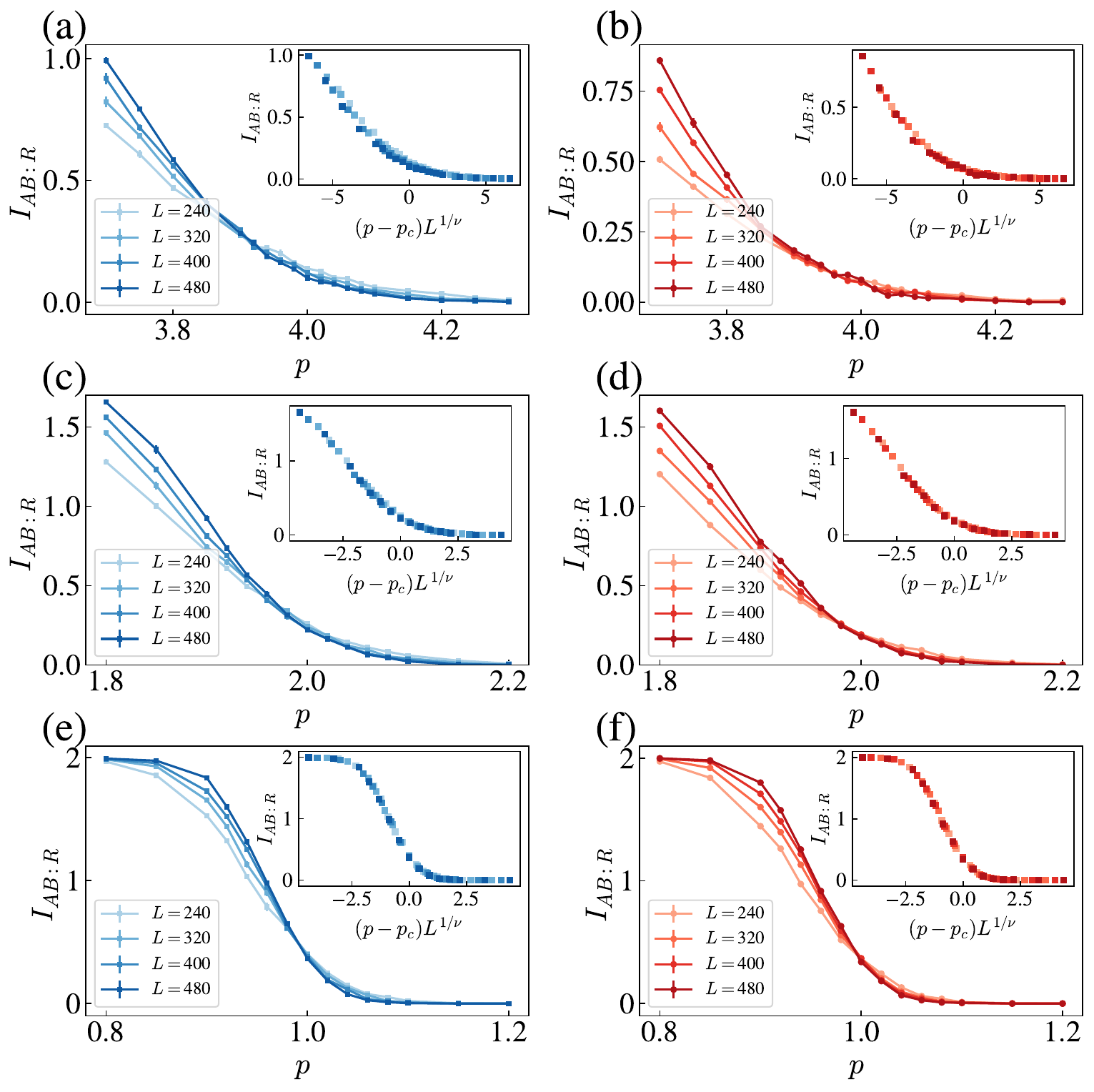}
\caption{The probability of reset channels is $q=p/L$ and the probability pf measurements is $p_{m}=0$. $T_{\text{scr}}=0$ in the left panel and $T_{\text{scr}}=L$ in the right panel. (a-b), (c-d), and (e-f) show the mutual information $I_{AB:R}$ with $T/L=1/4, \ 1/2, \ 1$ respectively. Insets show the data collapse with $p_{c}=L/T$ and $\nu=2$.}
\label{fig:Reset_pm0.0_Coding_diffz}
\end{figure}

\begin{figure}[ht]
\centering
\includegraphics[width=0.45\textwidth, keepaspectratio]{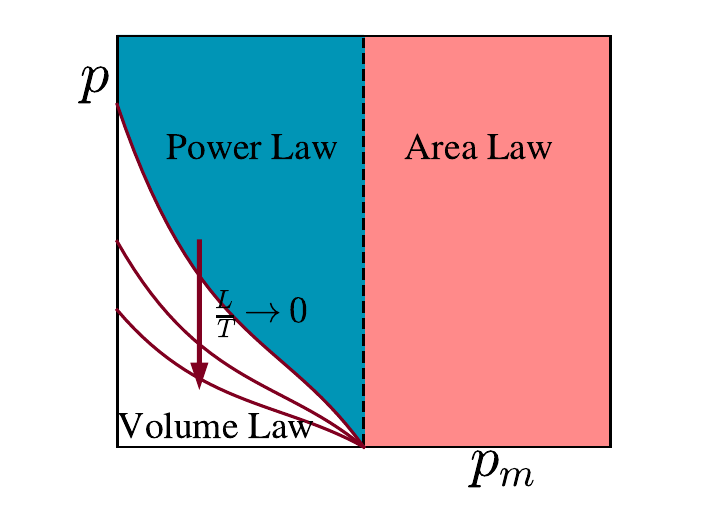}
\caption{Schematic phase diagram: $p_{c}$ decreases with the ratio $L/T$ decreases.}
\label{fig:phase_diagram_2}
\end{figure}

As discussed in the main text and shown in Fig. \ref{fig:Reset_pm0.0_Coding_diffz}, the critical probability of noises will decrease as the ratio $L/T$ decreases. 
Besides the phase diagram with $T/L=4$ in the main text, we also show a schematic phase diagram with varying ratio $L/T$ in Fig. \ref{fig:phase_diagram_2}. 
In the limit $L/T \rightarrow 0$, the noise-induced coding phase transition, as well as the noise-induced entanglement phase transition, disappear.
From the perspective of coding transition, it is consistent with the fact that the encoded information is ultimately destroyed by the quantum noises.

\section{Numerical results for MIPT in the presence of quantum noises}
\begin{figure}[ht]
\centering
\includegraphics[width=0.49\textwidth, keepaspectratio]{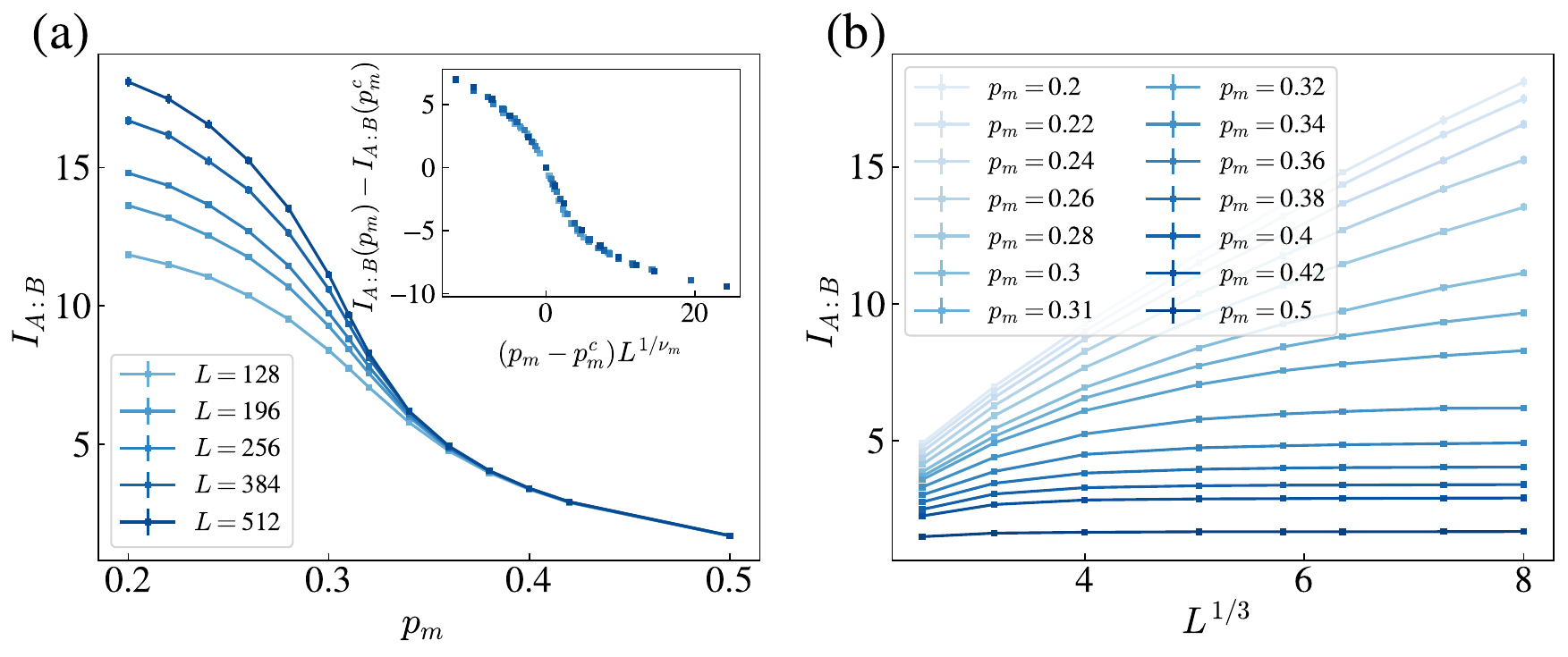}
\caption{The probability of reset channels is $q=0.252/L$ and $T=4L$. (a) $I_{A:B}$ vs $p_{m}$. Inset shows the data collapse with $p_{m}=0.3$ and $\nu_{m}=1.3$. (b) $I_{A:B}$ vs $L^{1/3}$. As the measurement probability increases, there is an entanglement phase transition from power law to area law.}
\label{fig:mipt_q0.252}
\end{figure}

In the section, we show more numerical results of measurement-induced entanglement phase transition in the presence of quantum noises with scaling exponent $\alpha=1$. As shown in Fig. \ref{fig:mipt_q0.252}, there is a measurement-induced entanglement phase transition from power law to area law with the increases of measurement probabilities. The critical probability of measurements $p_{m}^{c}$ and critical exponent $\nu_{m}$ are consistent with those in MIPTs without quantum noises. We note that in the presence of quantum noises at spatial boundaries, the critical probability is also the same as that without quantum noises but the critical exponent changes which may be caused by the limited system sizes~\cite{PhysRevLett.129.080501}.

\section{Numerical results for noise-induced complexity transition in random circuit sampling}
\label{sec:complexity}
\begin{figure}[ht]
\centering
\includegraphics[width=0.4\textwidth, keepaspectratio]{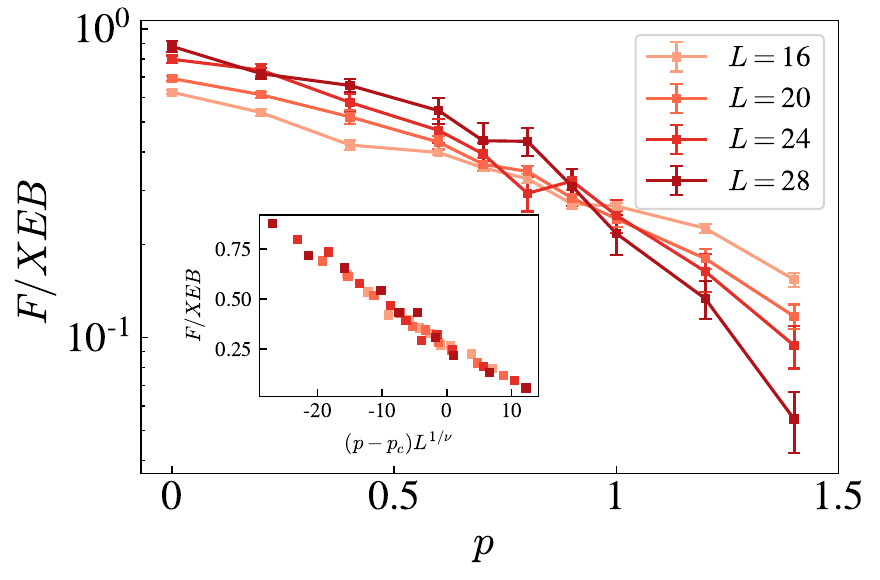}
\caption{The probability of quantum noises is $q=p/L$. The ratio of the averaged fidelity and the averaged XEB vs noise probability $p$. There is a noise-induced complexity transition. The inset shows the data collapse with $p_{c} \approx 0.96$ and $\nu \approx 1$.}

\label{fig:reset_XEB}
\end{figure}

\begin{figure}[ht]
\centering
\includegraphics[width=0.49\textwidth, keepaspectratio]{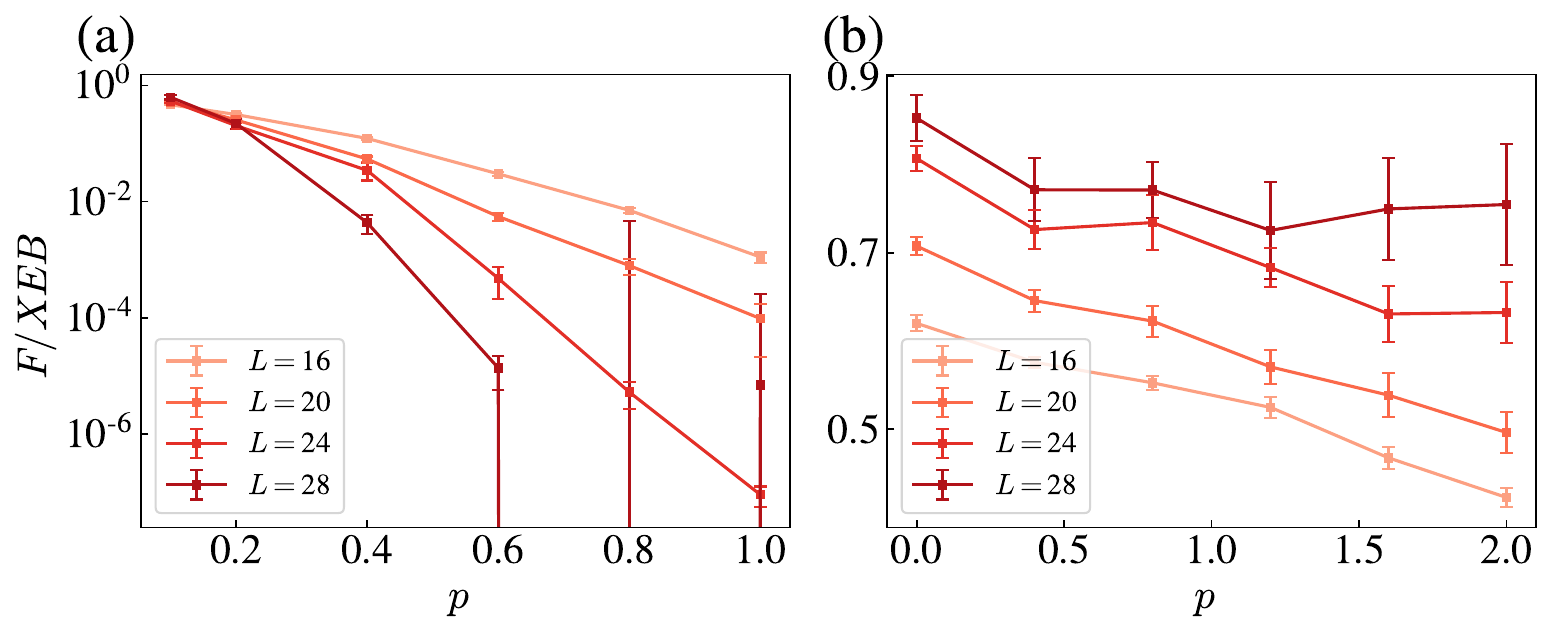}
\caption{Similar to cases of noise-induced entanglement and coding transitions, the complexity transition only exists with $\alpha=1$. (a)(b) show the ratio of the fidelity and XEB vs noise probability $p$ with scaling exponents $\alpha=0.5$ and $\alpha=1.5$.}
\label{fig:reset_XEB_diffscale}
\end{figure}

In addition to the noise-induced entanglement and coding transitions, there is also a noise-induced computational complexity transition in random circuit sampling~\cite{NIPT1_arxiv, NIPT2_arxiv, PhysRevA.109.042414}. When the quantum noise is strong, the wavefunction of the system can be approximately represented by multiple uncorrelated subsystems. This makes the quantum system vulnerable to spoofing by classical algorithms that only represent a part of the system. However, when the quantum noise is sufficiently weak, correlations span the entire system restoring its computational complexity. 
We demonstrate this
transition numerically by the crossing of the ratio of the fidelity and the linear cross-entropy benchmarking (XEB), which is defined as
\begin{eqnarray}
    XEB = 2^{L} \sum_{s} p(s)q(s) -1,
\end{eqnarray}
where $p(s)$ and $q(s)$ are the distribution probabilities of bitstring $s$ of the final state of a given trajectory without and with quantum noises respectively. The noise-induced computational complexity transition is illustrated in Fig. \ref{fig:reset_XEB} with noise probability $q=p/L$. When $\alpha \neq 1$, this complexity transition also disappears, see Fig. \ref{fig:reset_XEB_diffscale} for more details.

%

\end{document}